\documentclass[prb,twocolumn,showpacs,amsmath,amssymb,superscriptaddress,floatfix]{revtex4}

\usepackage[english]{babel}
\usepackage{amsfonts}
\usepackage{graphicx}
\usepackage{times}

\begin{document}

\title{
Emergence of states in the phonon spectral function of the Holstein polaron below and above the one-phonon continuum
}

\author{Lev Vidmar}
\affiliation{J. Stefan Institute, 1000 Ljubljana, Slovenia}

\author{Janez \surname{Bon\v ca}}
\affiliation{J. Stefan Institute, 1000 Ljubljana, Slovenia}
\affiliation{Faculty of Mathematics and Physics, University of Ljubljana, 1000
Ljubljana, Slovenia}

\author{Stuart A. Trugman}
\affiliation{Theoretical Division, Los Alamos National Laboratory, Los Alamos, New Mexico 87545, USA}




\date{\today}
\begin{abstract}
We investigate the low-energy properties of the Holstein polaron through calculation of the $q$-dependent phonon spectral function using an improved exact-diagonalization technique, defined over a variational Hilbert space. We perform a comprehensive study of the low-energy excitations of the polaron. 
Beside the energy range, where the additional phonon excitation is unbound, we observe separate coherent peaks which correspond to bound and antibound states of a polaron and additional phonon quanta. These novel states can be observed for intermediate and strong electron-phonon coupling strengths, as well as below and above the unbound one-phonon excitation spectrum.
A detailed investigation of their properties is presented. 
We find good agreement between the phonon spectral function obtained from the first-order strong-coupling perturbation theory and numerical results. 
\end{abstract}

 \maketitle

\section{Introduction}

Electron-phonon coupling represents one of the fundamental mechanisms that determines thermodynamic as well as transport properties of solids. Polaron formation is a phenomenon where a single charge carrier changes its state by absorbing or emitting phonon quanta. Its importance has been identified in many novel materials~\cite{alexandrov07}, such as high-temperature superconductors~\cite{lanzara01,gunnarsson08,mishchenko10}, colossal magnetoresistance materials~\cite{edwards02}, molecular crystals and fullerenes~\cite{gunnarsson97}. In the case where the electron couples to an optical branch of lattice vibrations, a widely accepted approximation is used where the electron-phonon (e-ph) coupling strength and the phonon energy are considered momentum independent. This leads to a Holstein molecular crystal model, which is one of the most fundamental models in solid state physics.
Many analytical and numerical methods have provided a fair understanding of its ground state properties, i.e., the formation of a polaron~\cite{osor3,alexandrov95,wellein97,wellein98,romero98,jeckelmann98}. 
Its behavior, particularly in the intermediate-coupling (IC) and strong-coupling (SC) regime, strongly depends on the strength of the phonon energy $\omega_0$.
In the antiadiabatic regime where the Lang-Firsov transformation~\cite{lang1} in the SC limit provides a comprehensive picture of the polaron, most approximations usually manage to calculate the system properties qualitatively correctly,
while the treatment of adiabatic quantum phonons poses a more challenging task and is a matter of the recent scientific interest~\cite{osor4,marsiglio10,alvermann10}.

Above the polaron ground state energy $E^0$, several features of the excited states can be observed. In particular, we are interested in the low-energy excited states which do not belong to the one-phonon continuum of an unbound phonon starting at $E^0 + \omega_0$.
A while ago, the emergence of a state below the one-phonon continuum has been reported~\cite{cho,engelsberg}. In addition, a coherent state above the continuum has been found using dynamical coherent potential approximation~\cite{sumi}.
In the last decades, the investigation of these states spread to all parameter regimes of the Holstein model.
In the SC adiabatic approximation~\cite{kabanov93,kalosakas}, several states were observed below the continuum.
When moving to non-zero phonon energies, one such state was found in the SC approximation~\cite{gogolin} and three states in the IC regime using exact-diagonalization technique~\cite{alexandrov94}. Recently, an investigation of properties of the coherent states below the one-phonon continuum was performed by variational exact diagonalization~\cite{bonca1,osor1} and momentum-averaging (MA) approach~\cite{goodvin,berciu07a,berciu07b}. 

Regarding dynamical properties of the Holstein polaron, the analysis rarely focused on
the states outside the continuum. 
Spectral weight was assigned to these states in the calculation of one-electron spectral function using dynamical mean-field theory (DMFT)~\cite{ciuchi97} and MA approach~\cite{berciu07b}, optical conductivity using DMFT~\cite{fratini01}, and in the quantum Monte Carlo study of the one-electron spectral function in the Rashba-Pekar model~\cite{mishchenko02}.
Less emphasis has been devoted to the phonon spectral function.
The latter was studied in Ref.~\cite{loos}, where authors numerically and analytically calculated the $q$-dependent phonon spectral function for various phonon energies and for all relevant e-ph coupling regimes.
In the low-lying excitation spectrum they find a mirror peak, which was supposed to represent a state lying an energy $\omega_0$ above the polaron peak. They did not focus on the description of the excited states below the one-phonon continuum.
On the other hand, the author of Ref.~\cite{osor2} studied lattice correlation functions of few excited states below the continuum in the IC regime, representing the spectral weight of the phonon spectral function. Calculating these correlations, the author showed that these states contain a non-vanishing spectral weight and should contribute to a formation of coherent excited bands.
Nevertheless, the features of the spectrum at and above $\omega_0$ were not studied.
Taking this into account, we argue that a comprehensive study of the low-lying features of the Holstein polaron phonon spectral function is still missing. In particular, the emergence of the states below and above the one-phonon continuum, denoted here as the novel states, needs to be characterized.

In this paper we study the low-energy spectrum of the Holstein polaron model, reflected through calculation of the $q$-dependent phonon spectral function. 
Using an efficient exact-diagonalization technique defined over a variational Hilbert space~\cite{bonca1}, we are able to investigate a polaron band as well as a non-dispersive continuum band starting at $\omega_0$ above the polaron ground state.
This band represents the states where the additional phonon excitation is not bound to the polaron.
In addition, the emergence of novel states with the energy below and above the one-phonon continuum is studied. These states are denoted in the text as bound and antibound states, respectively. 
Calculation of the static correlation function in these states shows that the weight of the extra phonon excitations decreases exponentially with the distance from the polaron.
For different values of the model parameters, a particular emphasis is given to the onset of the antibound state, not studied before in the literature.
The emergence of one bound and antibound peak in the phonon spectral function is studied within the framework of the first-order strong-coupling perturbation theory. For a certain range of the model parameters, this analytical calculation provides results that are in good agreement with the numerical solution.

The paper is organized as follows. In Sec. $\mbox{II.}$ we introduce the model and the numerical method. In Sec. $\mbox{III.}$ we show the numerical results, while in Sec. $\mbox{IV.}$ we compare the results with the first-order strong-coupling perturbation theory. A summary is given in Sec. $\mbox{V.}$

\section{Model and Numerical Method}
We start by writing the one-dimensional spinless Holstein model as
\begin{equation} \label{hol}
\vspace*{-0.0cm}
H = -t\sum_{\langle {i,j}\rangle}(c^\dagger_{i} c_{j} +\mathrm{H.c.})
 + \tilde{g} \sum_{i} n_{i} (a_{i}^\dagger + a_{i}) +
\omega_0\sum_{i} a_{i}^\dagger  a_{i},
\end{equation}
where $c^\dagger_{i}$ and $a^\dagger_{i}$ are electron and phonon creation operators at site $i$, respectively, and $n_{i} = c^\dagger_{i}c_{i}$ electron density at site $i$ (with $n_i=0,1$). The total number of electrons in the system is equal to $1$. 
$\omega_0$ denotes a dispersionless optical phonon energy and $t$ nearest-neighbor hopping amplitude (we set $t=1$ in Sec. $\mbox{I.-III.}$).
The e-ph coupling strength is denoted as $\tilde{g}$ and will be in the following replaced by a dimensionless coupling $g = \tilde{g}/\omega_0$.

The phonon Green function, that corresponds to the displacement-displacement correlation function, is defined as
\begin{equation} \label{D}
 D(q,t_1-t_2) = -i\langle \psi^0 \vert \hat{T} [ \hat{x}_q(t_1)\hat{x}_q^\dagger(t_2) ] \vert \psi^0 \rangle,
\end{equation}
where $\hat{x}_q = a_q + a_{-q}^\dagger$ and $\hat{T}$ the time ordering operator. Applying the Fourier transform and calculating the spectral representation $B(q,\omega)=- \frac{1}{\pi} \mbox{Im}D^R(q,\omega)$, where $\omega>0$ and $D^R(q,\omega)$ the retarded Green function, one gets the corresponding phonon spectral function
\begin{equation} \label{B}
 B_q(\omega) = \sum_n \vert\langle \psi_q^n \vert \hat{x}_{-q} \vert \psi^0 \rangle\vert^2 \delta(\omega - (E_q^n-E^0)),
\end{equation}
with $E^0$ and $\psi^0$ denoting the polaron ground state energy and wave function, respectively, while the sum runs over all excited states.
The ground state wave function $\psi^0$ has momentum $k=0$, and $q$ may be nonzero.

We used an improved numerical method, originally introduced in Ref.~\cite{bonca1}, which constructs the variational Hilbert space (VHS) starting from the single-electron Bloch state $c_{\bf k}^{\dagger} \vert \emptyset{\rangle}$ on infinite lattice. The VHS is then generated by applying the off-diagonal terms of Hamiltonian (\ref{hol}) to the starting state
\begin{equation}
\left \{ \vert \phi_{{\bf k},l}^{(N_h,M)}{\rangle}\right \} = \left ( H_{\rm kin} + 
H_{\rm g}^M\right )^{N_h} c_{\bf k}^{\dagger} \vert \emptyset{\rangle},
\end{equation}
where $H_{\rm kin}$ and $H_{\rm g}$ corresponds to the first and the second term of the Hamiltonian in Eq.~\ref{hol}, respectively. The set of basis states is determined by parameters $N_h$ and $M$, where $N_h-1$ is the maximum distance between the electron and the phonon quanta and $N_h*M$ is the maximum number of phonon quanta contained in the Hilbert space. The parameter $M>1$ first introduced in Ref.~\cite{bonca3}, is chosen to ensure good convergence for the strong e-ph coupling in the adiabatic regime where the low-energy states contain multiple phonon excitations. On the other hand, large values of $N_h$ provide a fair description of the polaron states also in the weak coupling regime, where the spacial extent of a lattice deformation around the electron is large. We have used $N_h\geq8$ and $M\geq4$ that lead to converged results in all parameter regimes. 
Note that for any Hilbert space in our calculations the number of sites is infinite.
The convergence towards thermodynamic limit is then achieved when the number of phonon states for a given electron location is sufficiently increased.
Once the VHS is generated, the Holstein Hamiltonian is diagonalized using standard Lanczos procedure, where translational symmetry is explicitly taken into account.

\section{Results}

\begin{figure}[!tbh]
\includegraphics[width=1.0\columnwidth]{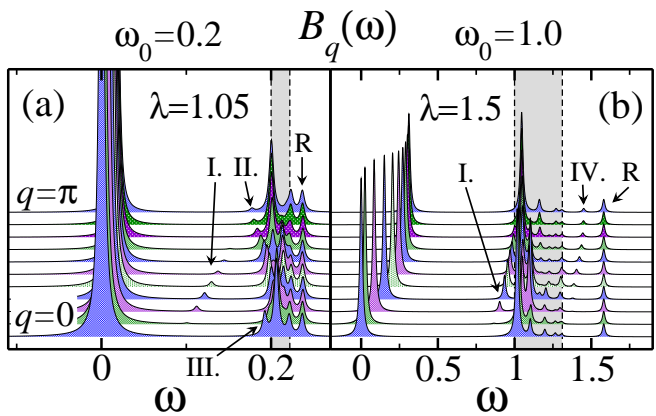}
\caption{(Color online) (a) $q$-dependent phonon spectral function $B_q(\omega)$ for $q=0,...\pi$ and $\omega_0=0.2$, $\lambda=\tilde{g}^2/2t\omega_0=1.05$. Only the polaron peak and one-phonon excitation spectrum is shown. We have used an artificial damping of $\epsilon=0.003$. The Roman numerals $\mbox{I.}$, $\mbox{II.}$, $\mbox{III.}$ denote the three bound states.
(b) $B_q(\omega)$ for $\omega_0=1.0$ and $\lambda=1.5$. Here only the lowest bound state (Roman numeral $\mbox{I.}$) can be observed. An artificial damping of $\epsilon=0.01$ was used.
The Roman numeral $\mbox{IV.}$ denotes the antibound peak, where the one-phonon excitation has an energy higher than $\omega_0+W$.
The non-dispersive peak (denoted by R) corresponding to one-phonon excitation repelled from the polaron, is located slightly above $\omega_0+W$. This peak is described in more detail in Fig.~\ref{fig2} and Fig.~\ref{fig4} and further in the text.
The grey areas correspond in (a) and (b) to a continuum of states in the interval $[\omega_0,\omega_0+W]$. 
We used the Hilbert space obtained for $N_h=8$, $M=11$ that lead to $N_{\mbox{st}}=5.9\times10^5$.
}\label{fig1}
\end{figure}

To start the calculation of the phonon spectral function, defined in Eq.~\ref{B}, we first focus on the lowest-energy peak that represents the polaron ground state. Its spectral weight is proportional to the electron density $1/N$ and increases when approaching the SC limit as $g^2$. In the adiabatic regime $\omega_0 \ll 1$ the cross-over from large to small polaron occurs for $\lambda=\epsilon_p/2t\geq 1$, where $\epsilon_p=g^2\omega_0$ is the polaron energy at $t=0$ and $\lambda$ a dimensionless e-ph coupling. In Fig.~\ref{fig1}(a) the $q$-dependent phonon spectral function is plotted for $\omega_0=0.2$ and $\lambda=1.05$. 
At this e-ph coupling strength the bandwidth of the polaron $W$ is reduced to $W=\zeta W_0 = W_0/10$, where $W_0$ is the polaron bandwidth in the weak-coupling (WC) limit ($W_0=\omega_0$ if $\omega_0<4$ and $W_0=4$ otherwise) and $\zeta$ a dimensionless parameter. 
When $\omega_0$ is increased away from the adiabatic regime, the cross-over to SC regime is less abrupt. In Fig.~\ref{fig1}(b) it can be seen for $\omega_0=1.0$ and $\lambda=1.5$ that the polaron dispersion is still substantial and $\zeta=0.31$.

We now turn to the investigation of the low-lying excited states, representing the main focus of the paper. 
Naively, the low-lying excited states should consist of a polaron at a particular $k$-point and an additional unbound phonon excitation with momentum $Q$, leading to an arbitrary total momentum $q=k+Q$ of the composite excited state.  
This implies that in the thermodynamic limit of the phonon spectral function, the unbound excited states would have a non-vanishing spectral weight in the interval $[\omega_0,\omega_0+W]$. In Fig.~\ref{fig1}, this continuum of states starting at $\omega_0$ is shaded and its width is given by the bandwidth $W$ of the polaron peak. Due to the numerical restrictions of the variational Hilbert space, the calculations do not show a considerable spectral weight throughout the whole continuum band, even though its approximate width is well reproduced. When increasing the Hilbert space, the density of states inside the continuum band grows. The effects of the finite-size Hilbert space will be discussed later.

Beside the unbound one-phonon excitations, it is well known that in the IC regime of the Holstein model a bound state of a polaron and additional phonon excitations appears, where the additional phonon excitations are bound to the polaron and the energy of such composite state is less than $\omega_0$ above the ground state.
This phenomenon is sometimes denoted as softening. 
It has been shown~\cite{bonca1,osor1} that the average value of phonon quanta in the bound state $N_{ph}^1=\langle\sum_i a_i^\dagger a_i\rangle$ can not be related to the corresponding value of the ground state as $N_{ph}^1=N_{ph}^0+1$. The latter holds true only for the unbound one-phonon excitations.
In general, there are two bound states with different symmetry of the wave functions. 
In Fig.~\ref{fig1}(a) for $\omega_0=0.2$ and $\lambda=1.05$, the lowest bound state denoted by the Roman numeral $\mbox{I.}$ has a symmetric wave function at $q=0$ and an antisymmetric wave function at $q=\pi$, while the second bound state (the Roman numeral $\mbox{II.}$) is symmetric at $q=\pi$. Their spectral weight in $B_q(\omega)$ is approximately proportional to electron density $1/N$ and vanishes at $q=0$ state~\cite{osor2}. 
In the case of $\lambda=1.05$, an additional excited state (the Roman numeral $\mbox{III.}$) appears below the phonon threshold $\omega_0$. This state, however, has a non-vanishing spectral weight also at $q=0$.
The emergence of the bound peaks in $B_q(\omega)$ is best resolved in the IC regime as shown in Fig.~\ref{fig1}, since with the increasing of e-ph coupling towards the SC regime, their positions approach the phonon threshold $\omega_0$.
On the other hand, when the phonon energy is increased (the case for $\omega_0=1.0$ is shown in Fig.~\ref{fig1}(b)) the positions of the bound peaks shift as well closer to the bare phonon energy. In this case only the lowest bound state (the Roman numeral $\mbox{I.}$) can be resolved from the spectrum.

While the non-zero spectral weight in $B_q(\omega)$ for the peaks below the phonon threshold was suggested in previous works~\cite{osor2}, we find a novel coherent peak located above the continuum band of unbound one-phonon excitations. This peak is denoted by the Roman numeral $\mbox{IV.}$ in Fig.~\ref{fig1}(b). 
In analogy to the bound states, this peak represents a state where the polaron and additional phonon excitations are antibound, i. e.,  its energy is above the upper edge of the continuum band of the unbound one-phonon excitations, $E_q^a>\omega_0+W$.
Note that for $\omega_0=0.2$, as seen in Fig.~\ref{fig1}(a) no antibound peak can be observed as its spectral weight is vanishingly small.
One of the first indications that the energy spectrum of the Holstein polaron can contain a state above the one-phonon continuum for the non-adiabatic values of the phonon energy, was reported many years ago in the calculation based on dynamical coherent potential approximation~\cite{sumi}.
Recently, coherent states above the one-phonon continuum were observed in the study of one-electron spectral function using DMFT~\cite{ciuchi97} and MA approach~\cite{berciu07b}.

\begin{figure}[!htb]
\includegraphics[width=1.0\columnwidth,clip]{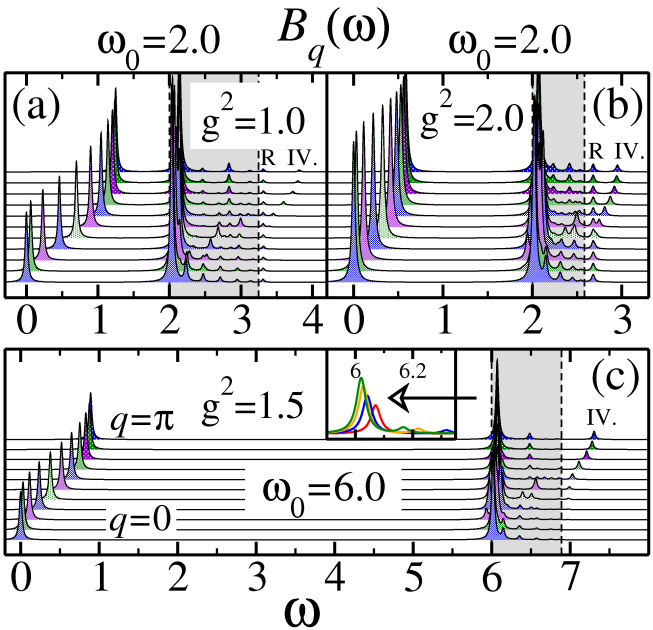}
\caption{(Color online) (a), (b) $q$-dependent phonon spectral function $B_q(\omega)$ for $g^2=1.0$ (where $\zeta=0.62$) and $g^2=2.0$ ($\zeta=0.29$), respectively, and $\omega_0=2.0$. (c) $q$-dependent phonon spectral function $B_q(\omega)$ for $g^2=1.5$ ($\zeta=0.22$) and $\omega_0=6.0$. The grey areas correspond to a continuum of states in the interval $[\omega_0,\omega_0+W]$. An artificial damping of $\epsilon=0.02$ was used in all cases. The Hilbert space used in the calculation was obtained for $N_h=11$, $M=8$. 
The inset of (c) shows $B_q(\omega)$ around $\omega_0\gtrsim6.0$ at $q=\pi$ for different system sizes. The lowest peak (red curve) corresponds to a Hilbert space $(N_h,M)=(8,11)$, $N_{\mbox{st}}=5.9\times10^5$. As the weight of the peak increases and its position shifts to lower energies, Hilbert space is changed to $(10,7)$, $(12,7)$ and $(14,4)$, which implies $N_{\mbox{st}}=1.8\times10^6,2.5\times10^7$ and $1.8\times10^7$, respectively.
}\label{fig2}
\end{figure}

The antibound peak is more pronounced for the higher values of $\omega_0$. In Fig.~\ref{fig2}(a), (b) and Fig.~\ref{fig2}(c) we plot the $q$-dependent phonon spectral function for $\omega_0=2.0$ and $6.0$, respectively. As the strength of the dimensionless e-ph coupling $g^2$ approaches the SC limit, the spectral weight of the antibound peak increases.
The energy of its state relative to the ground state $E_q^a$ is the highest at $q=\pi$, however its dispersion is not proportional to the polaron dispersion. When the momentum is decreased, the energy decreases, but always remains above the continuum band. 
Its spectral weight vanishes at the $q=0$ point, where the symmetry of the antibound wave function is antisymmetric. On the other hand, the wave function is symmetric at $q=\pi$.
Note that in all cases under investigation when $\omega_0 > 1$, we use the dimensionless e-ph coupling $g=\tilde{g}/\omega_0$. In the antiadiabatic limit $\omega_0\gg 1$, the condition $g\geqslant 1$ determines the cross-over from large to small polaron.

The phonon spectral function calculated in Ref.~\cite{loos} using the kernel polynomial method can be compared to our results. We find a similarity between the antibound peak as denoted in our paper and the mirror peak as defined in the latter reference. Even though the authors of Ref.~\cite{loos} did not examine the structure of this peak, we may speculate that these two peaks represent the same state. In the second part of this section, we shall further investigate the properties of the antibound state through calculation of the e-ph correlation function (Fig.~\ref{fig4}) and its evolution when the e-ph coupling increases (Fig.~\ref{fig5} and ~\ref{fig6}).

In addition to the bound and antibound peaks, described above, another feature of the phonon spectral function can be observed in Fig.~\ref{fig1} and Fig.~\ref{fig2}(a) and (b), denoted by the letter R. This is a non-dispersive peak located slightly above $\omega_0+W$ with the spectral weight comparable to the peaks due to bound and antibound phonon excitations. However, due to the non-dispersive nature of the peak, its origin should be different from the latter peaks.
Even though it seems from Fig.~\ref{fig1} and Fig.~\ref{fig2} that its spectral weight is not sensitive to the value of $\omega_0$, it suddenly vanishes as $\omega_0>4.0$, as demonstrated in Fig.~\ref{fig2}(c) for $\omega_0=6.0$. We will show later in the calculation of the e-ph correlation function (Eq.~\ref{ephnum}) that this state is formed by a polaron and an additional phonon excitation, where the latter is repelled from the polaron in real space. 
We shall argue as well that the position of this peak approaches $\omega_0+W$ from above as the size of the Hilbert space increases and in the thermodynamic limit merges with the continuum.
In contrast, spectral weight and positions of peaks $\mbox{I.}$ to $\mbox{IV.}$ are well converged in the thermodynamic limit.

The energy of the polaron ground state calculated by our numerical method is variational in the thermodynamic limit as $N_h$ $\to$ $\infty$ and can be calculated to a great accuracy~\cite{bonca4}. This is also reflected in the calculation of the polaron energy in different $q$-points, where its peak position is not sensitive to different sizes of Hilbert space (not shown in the paper). For the unbound one-phonon excitations, where the additional phonon excitation is not correlated to the polaron position and can be located anywhere on the lattice, the energy convergence is slower. The dominant contribution of the unbound phonon excitations in the phonon spectral function comes from the peak at $\omega_0$ above the polaron ground state. In our calculations the position of this peak is located at $\omega_0+\epsilon$ for a finite $N_h$, where $\epsilon$ decreases when $N_h$ is increased. For a fixed system, $\epsilon$ is the largest in the WC limit. 
In the current work, our main focus is the behavior of the phonon spectral function in the IC regime. 
Generation of a sufficient number of phonon excitations in the vicinity of the charge carrier in this regime, starts to play a more crucial role on the system properties than the maximal distance between polaron and the extra phonon excitation.
While the first condition is well controlled by the value of the parameter $M$ in the generation of VHS, the latter is determined by the parameter $N_h$. We use this subtle interplay between the values of these two parameters generating the Hilbert space to achieve the convergence of both unbound, bound and antibound low-energy states in the phonon spectral function. The inset of Fig.~\ref{fig2}(d) show the position of the $\omega_0$-peak for $q=\pi$, $g^2=1.5$ and $\omega_0=6.0$. As $N_h$ increases from $N_h=8$ to $14$, the peak indeed approaches the bare phonon energy and gains the spectral weight. We have thus shown that for various sizes of the Hilbert space included in our calculation, the $\omega_0$-peak is well reproduced.

\begin{figure}[htb]
\includegraphics[width=1.0\columnwidth,clip]{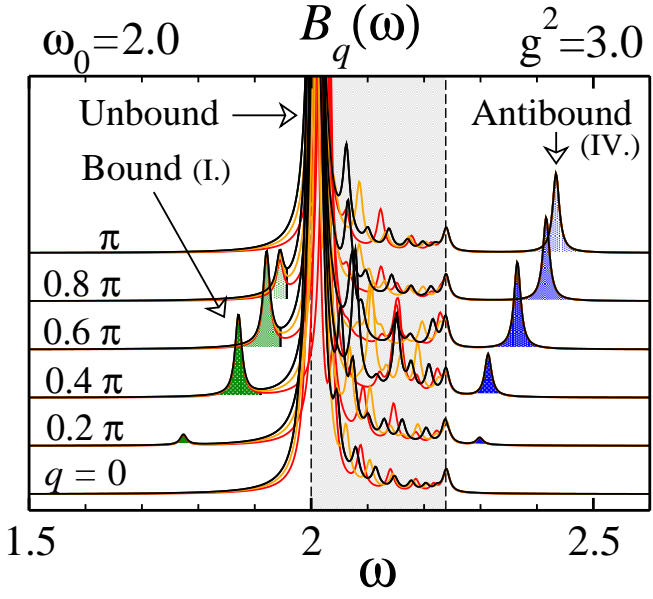}
\caption{(Color online) $B_q(\omega)$ for $g^2=3.0$ and $\omega_0=2.0$ ($\zeta=0.12$). The polaron peak is not shown and the repulsive peak is removed from the figure. The grey area corresponds to the continuum band of unbound states in the interval $[\omega_0,\omega_0+W]$. An artificial damping of $\epsilon=0.007$ was used. The three curves were obtained using the following parameters $(N_h,M)$: $(10,4)$ - red, $(12,4)$ - yellow and $(14,4)$ - black, that lead to the following number of basis states: $N_{\mbox{st}}=2.2\times10^5,2.0\times10^6$ and $1.8\times10^7$. The peaks below and above the grey area, where the curves are nearly overlapping, correspond to the lowest bound state (green) and to the antibound state (blue), respectively.
}\label{fig3}
\end{figure}

It is crucial to show that not only the $\omega_0$-peak, but the bound and antibound peaks as well as the width of the continuum band converge well within the framework of our numerical method. 
This can be seen from Fig.~\ref{fig3}, where the low-energy excitation spectrum of $B_q(\omega)$ is plotted for $\omega_0=2.0$ and $g^2=3.0$. Results are obtained for different system sizes $N_h=10,12$ and $14$ while the parameter $M$ is fixed to $4$. There are two important features in this figure. The first one concerns the continuum of the unbound phonon excitations in the interval $[\omega_0,\omega_0+W]$. The density of the states in this interval depends on the maximal number of allowed phonon quanta in our calculation as well as on the maximal allowed distance between the phonon quantum and the polaron. As the parameter $N_h$ is increased, the unbound states become denser, leading in the thermodynamic limit to the continuum of states with the non-zero spectral weight. 
On the other hand, the lowest bound peak and the antibound peak are located outside the continuum. 
Since in the case of the bound and antibound states the additional phonon quanta are attached to the polaron,
their position and weight in the phonon spectral function should be less sensitive to the finite Hilbert space. Indeed, the three nearly overlapping curves for different Hilbert spaces in Fig.~\ref{fig3} indicate that these peaks are well converged states, clearly separated from the continuum of states. This is not surprising since in this case, as we shall see in Fig.~\ref{fig4}, the amplitude of the extra phonon excitations decrease exponentially with the distance from the polaron and can be described efficiently within the VHS.

In order to investigate the structure of the observed peaks of the $B_q(\omega)$ in more detail, we calculate the static electron-phonon correlation function
\begin{equation} \label{ephnum}
 \gamma_q^n(i-j) = \langle \psi_q^n \vert c_{i}^\dagger c_{i} a_{j}^\dagger a_{j} \vert \psi_q^n \rangle,
\end{equation}
which represents the distribution of the number of excited phonon quanta in the vicinity of the electron. 
The strength of the numerical method is reflected in the fact that $\gamma_q^n(i-j)$ can be calculated for $(i-j)>10$ in any state $q$ as well as for any low-lying excited state $n$ obtained by the Lanczos diagonalization. 

\begin{figure}[!htb]
\includegraphics[width=1.0\columnwidth,clip]{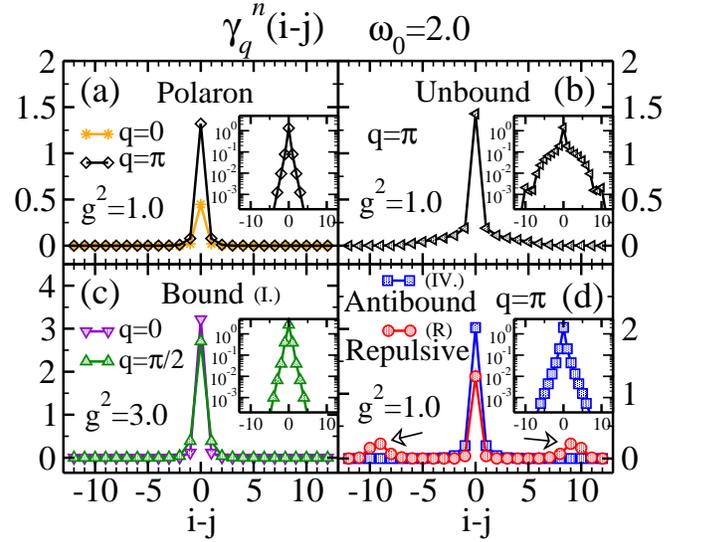}
\caption{(Color online) Electron-phonon correlation number $\gamma_q^n(i-j)$ for the ground state and various excited states at $\omega_0=2.0$. 
(a) Polaron ground state at $q=0$ and $q=\pi$ for $g^2=1.0$. 
(b) An excited state with momentum $q=\pi$ and $g^2=1.0$, which lies an energy $\omega_0$ above the state in the inset of (a). 
(c) The lowest bound state at $q=0$ and $q=\pi/2$ for $g^2=3.0$. 
(d) Antibound and repulsive state at $q=\pi$ and $g^2=1.0$. Note that the integral for the repulsive state over the two humps marked by arrows, is equal to $1.0$.
The insets show the correlation functions on a logarithmic scale.
The Hilbert space used in these calculations was obtained for $(N_h,M)=(12,7)$, $N_{\mbox{st}}=2.5\times10^7$.
}\label{fig4}
\end{figure}

Using $\gamma_q^n(i-j)$ of Eq.~(\ref{ephnum}), different properties of the excited states can be analyzed. For example, the unbound state can be distinguished from the bound and antibound state. In Fig.~\ref{fig4} we plot $\gamma_q^n(i-j)$ at $\omega_0=2.0$. We start with the polaron state at $g^2=1.0$, Fig.~\ref{fig4}(a). The ground state at the $q=0$ point has the minimum number of phonon quanta, located in the vicinity of the electron, while the number of phonon quanta increases when $q$ approaches the Brillouin-zone boundary. To show the spatial correlation of the electron and phonon position, the exponential decay of $\gamma_q^0(i-j)\vert_{q=\pi}$ is shown in the inset of Fig.~\ref{fig4}(a). 
The one-phonon excitation of the latter state is shown in Fig.~\ref{fig4}(b).
As seen from the inset of the figure, this state does not exhibit an exponential decay, i.e., the additional phonon excitation is unbound.

It is interesting to compare Fig.~\ref{fig4}(b) with (c) and (d), where we plot $\gamma_q^n(i-j)$ for the first bound (at $g^2=3.0$) and the antibound state (at $g^2=1.0$), respectively.
Note that in Fig.~\ref{fig3} the finite size investigation was performed for $\omega_0=2.0$ at $g^2=3.0$ due to the poor resolution of the bound peak for the lower values of the e-ph coupling (compare with Fig.~\ref{fig2}(a) and (b)).
Nevertheless, the antibound peak detaches at the $q=\pi$ point from the continuum already at $g^2\simeq0.77$.
In the inset of Fig.~\ref{fig4}(c) and (d), $\gamma_q^n(i-j)$ is shown on a logarithmic scale for $q=\pi/2$ (bound state) and $q=\pi$ (antibound state), respectively. 
For both cases the exponential decay confirms that in such composite excited states, the extra phonon excitations are bound to the polaron.

Other important information can be obtained from $\gamma_q^n(i-j)$. As already mentioned in the discussion of Fig.~\ref{fig1} and~\ref{fig2}, there is a non-dispersive peak in $B_q(\omega)$ denoted by R with a considerable spectral weight located above the continuum of unbound phonon excitations.
When increasing the size of the Hilbert space (not shown in the paper), its spectral weight does not scale to zero and remains a well defined quantity.
In Fig.~\ref{fig4}(d), the structure of this state can be resolved for $(N_h,M)=(12,7)$. Its striking features are two humps, well separated from the polaron peak. The integral over each of the humps gives $0.5$ for any momentum, which implies that an additional phonon excitation is repelled from the polaron. Consequently we call the non-dispersive peak obtained in Fig.~\ref{fig1} and~\ref{fig2}, the repulsive peak. 
Due to the repulsion between the polaron and an extra phonon excitation, its energy will always remain above $\omega_0+W$ in the finite Hilbert space calculations.
Thus the energy gap between this peak and the continuum band is an artifact of the method and scales to zero in the thermodynamic limit. 
Consequently, the repulsive peak would merge with the upper edge of the continuum band of the unbound states.

\begin{figure}[!htb]
\includegraphics[width=1.0\columnwidth,clip]{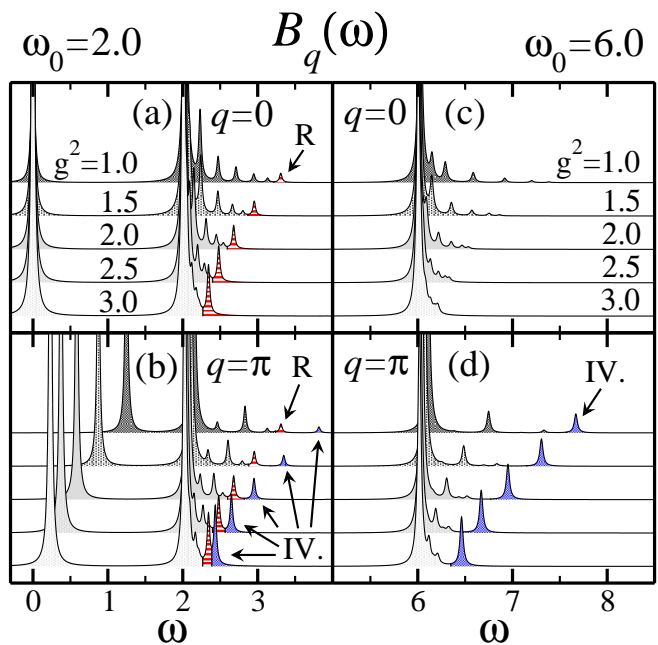}
\caption{(Color online) $B_q(\omega)$ for different values of e-ph coupling strength. (a) $q=0$ and (b) $q=\pi$, both at $\omega_0=2.0$. (c) $q=0$ and (d) $q=\pi$, both at $\omega_0=6.0$. For the latter phonon energy, the polaron peak is not shown. An artificial damping of $\epsilon=0.02$ was used. The Hilbert space was obtained for parameters $(N_h,M)=(8,11)$. The blue peak, denoted by the Roman numeral $\mbox{IV.}$, represents the antibound peak, while the red peak dashed by horizontal lines and denoted by R represents the repulsive peak.
}\label{fig5}
\end{figure}

By now, our aim has been to identify and introduce all the distinct features of the phonon spectral function and to prove their robustness to the finite-size Hilbert space. We would now like to get some general insight into the emergence of these novel states for different values of the phonon energy and the e-ph coupling strength.

In Fig.~\ref{fig5} we plot the phonon spectral function for a fixed $q$ and different values of $g^2$. We focus here mainly on the antibound and repulsive peak. In Fig.~\ref{fig5}(a) and (b), $B_q(\omega)$ is shown for $q=0$ and $q=\pi$, respectively, and $\omega_0=2.0$. At these values of parameters, the particular $q$-points do not show any spectral weight of the bound states. On the other hand, the repulsive peak denoted by R located slightly above $\omega_0+W$, is clearly visible and gains its spectral weight when the e-ph coupling is increased. 
The peaks in the continuum band of the unbound one-phonon excitations display a maximum at $\omega_0$, while their distribution and spectral weights decrease towards the upper edge $\omega_0+W$.
Nevertheless, due to the emergence of the repulsive state in our calculations,
we would anticipate an increase of the spectral weight at the upper edge of the continuum band.
When $\omega_0$ is increased to $6.0$, as seen in Fig.~\ref{fig5}(c) and (d), the repulsive state and correspondingly the nonmonotonous decay of the unbound excitation spectrum disappears.

The antibound peak observed in the same spectra for $q=\pi$ denoted by the Roman numeral $\mbox{IV.}$ is detached from the continuum of states and gains as well its spectral weight when the e-ph coupling is increased.
In addition, it is more pronounced for higher values of $\omega_0$ when calculated for the same values of the dimensionless e-ph coupling $g^2$. This should be compared to Fig.~\ref{fig1}(b) where $B_q(\omega)$ is plotted at $\omega_0=1.0$ and $\lambda=1.5$ (i.e., $g^2=3.0$). In this figure the antibound peak can be hardly observed. Note that there is the repulsive peak in Fig.~\ref{fig1}(b) located above the well converged antibound peak. As the system size $N_h$ is further increased, the repulsive peak moves to lower energies, as indicated before.
 
\begin{figure}[!htb]
\includegraphics[width=1.0\columnwidth,clip]{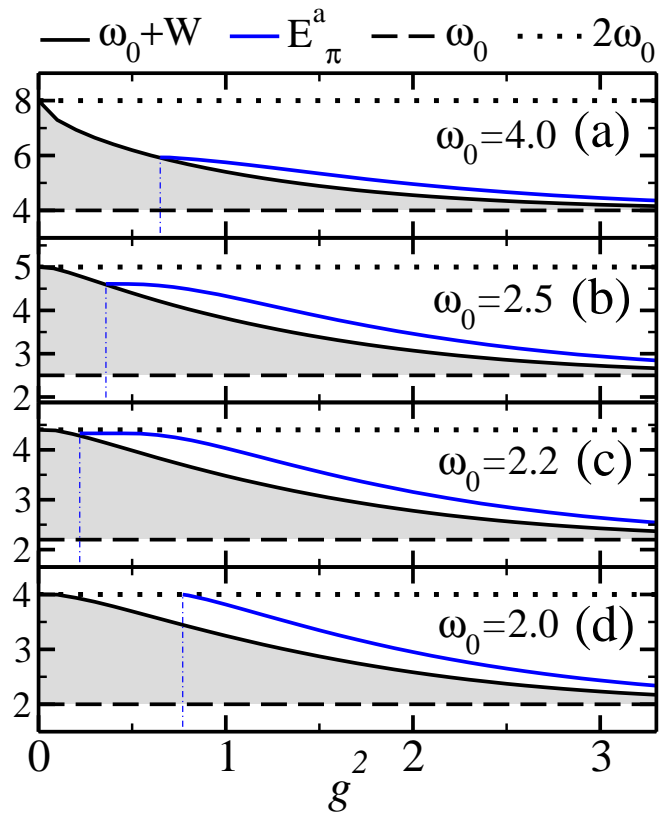}
\caption{(Color online) Energy positions of different states in the phonon spectral function above the $\omega_0$ threshold vs. the strength of the e-ph coupling. Dashed line marks the state at $\omega_0$ above the polaron ground state, while the dotted line marks the state at $2\omega_0$.
Bound states are not shown in the figure.
The black curve denotes the $\omega_0+W$ state and the blue curve the antibound state at $q=\pi$ with the energy $E^a_{\pi}$. Vertical dashed-dotted line marks the lowest value of $g^2$, at which the antibound peak can be observed at a particular $\omega_0$. The Hilbert space was obtained for parameters $(N_h,M)=(8,8)$ that lead to $N_{\mbox{st}}=2.0\times10^5$.
}\label{fig6}
\end{figure}

In general, there is no theoretical argument against emergence of additional coherent states above the $\omega_0 + W$ threshold in the energy spectrum of the Holstein polaron. In Sec. IV. we calculate the one-phonon excitation spectrum of the polaron in the SC antiadiabatic limit, as described within the Lang-Firsov picture. The first order perturbation theory predicts only one state above the $\omega_0 + W$ continuum in this limit, consistent with our numerical results. As seen from Fig.~\ref{fig5} and ~\ref{fig6}, this state, denoted throughout the paper as the antibound state, persist up to IC regime of e-ph coupling and moderate values of $\omega_0$. However, in the adiabatic limit $\omega_0 \ll 1$ where multiple excited coherent states exist below the $\omega_0$ continuum, it is likely that several coherent states emerge above the continuum as well. In terms of the phonon spectral function as shown in Fig.~\ref{fig1}(a), the latter states have vanishingly small spectral weight and are not discussed in the paper.

\begin{figure}[!tbh]
\includegraphics[width=1.0\columnwidth,clip]{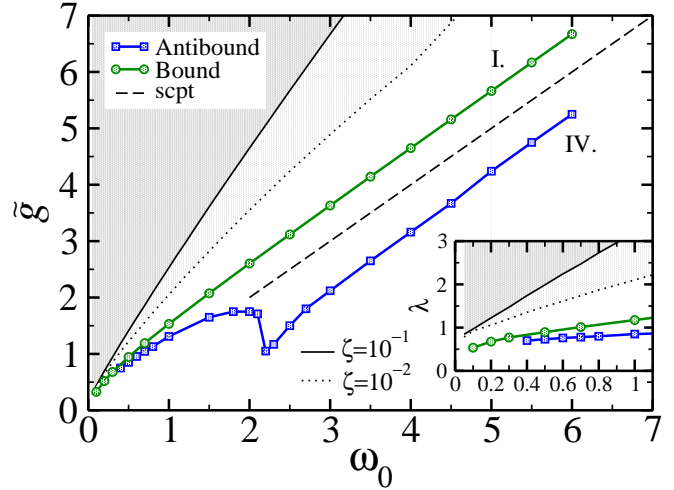}
\caption{(Color online) The minimum strength of the e-ph coupling for a fixed value of $\omega_0$, at which the bound or antibound state detach from the band of the unbound one-phonon excitations. The antibound state is plotted for $q=\pi$ and the bound state for $q=0$. 
Note that the e-ph coupling energy $\tilde{g}$ was introduced in Eq.~\ref{hol}.
The dashed line represents the case when $\tilde{g}=\omega_0$ (i.e., the result from the first-order scpt for both bound and antibound state), while the dotted and solid line correspond to the cases when the polaron bandwidth is reduced to $W=\zeta W_0$.
The inset shows the same results in terms of the dimensionless adiabatic parameter $\lambda$.
}\label{fig7}
\end{figure}

When addressing the emergence of the antibound state in $B_q(\omega)$, a question arises at which values of e-ph coupling the corresponding peak can be resolved from the spectrum. 
Since in the last part of this section (Fig.~\ref{fig7}) we show that the onset of this peak exhibit a non-monotonous behavior when the model parameters are changed, we need to get some more insight into the properties of the antibound state.

In Fig.~\ref{fig6} the energy positions relative to the polaron ground state energy $E^0$ for different states above the $\omega_0$ threshold are plotted vs. the strength of the e-ph coupling. The gray area corresponds to the continuum band with the upper edge $\omega_0 + W$ and the blue curve, i.e., the full line emerging out of the continuum, represents the antibound state with the energy $E^a_{\pi}$ at $q=\pi$.
A particular emphasis is given to the onset of the antibound state. A vertical dashed-dotted line marks the lowest value of $g^2$, for which this state can be observed.
It emerges out of the one-phonon continuum if $\omega_0 > 2.0$ (see Fig.~\ref{fig6}(a)-(c)), or out of higher-energy continuum if $\omega_0 \leq 2.0$ (see Fig.~\ref{fig6}(d)).



The above analysis shows that the antibound peak is not part of the continuum band and its energy is always higher than $\omega_0$ above the top of the polaron band. Thus we propose a slightly different interpretation of this peak than in Ref.~\cite{loos}. In our picture the peak corresponds to the state where the additional phonon excitations are attached to the polaron. This is qualitatively different from the states in the continuum band, where the extra phonon excitation is not bound to the polaron. In Sec. $\mbox{IV.}$ we will derive similar results from the calculation of $B_q(\omega)$ within the first-order strong-coupling perturbation theory.

To conclude the discussion we present a diagram in Fig.~\ref{fig7}. For a particular value of $\omega_0$, it shows the minimal strength of the e-ph coupling that the bound or antibound state would detach from the band of the unbound one-phonon excitations. For the antibound state, the minimum e-ph coupling is in the state $q=\pi$, while for the lowest bound state the minimum appears at $q=0$. Note that for the bound peak, its spectral weight is zero at $q=0$. Consequently, the corresponding curve in Fig.~\ref{fig7} represents just its lower boundary. It is in good agreement with Ref.~\cite{bonca1, goodvin}, where both analytical approximations as well as numerical calculations are compared.
As in Fig.~\ref{fig7} different regimes of $\omega_0$ and $\tilde{g}$ are included in the diagram, not all of the parameters are of physical interest. For a fixed $\omega_0$, a state at $\zeta=W/W_0=10^{-1}$ is represented by a dotted line and a state at $\zeta=10^{-2}$ by a solid line. Above the latter value, the polaron is self-trapped and can be in the non-adiabatic regime well described by the Holstein Hamiltonian in the atomic limit $t=0$.

As already noticed in Fig.~\ref{fig6}, where the vertical dashed-dotted lines mark the onset of the antibound peak in $B_q(\omega)$, the corresponding curve in Fig.~\ref{fig7} shows a non-monotonous behavior. It can be divided in two parts connected to each other at $\omega_0\approx2.0$.
While in the adiabatic regime the spectral weight of the antibound state is vanishingly small (see Fig.~\ref{fig1}) and its emergence begins at the e-ph coupling values comparable to that of the bound states, the most interesting behavior starts at $\omega_0\gtrsim 2.0$.
In particular, for $2.2<\omega_0<3.0$, the onset of the antibound peak occurs at $g^2<0.5$, i.e., for the values of e-ph coupling in the IC regime. The dashed curve in Fig.~\ref{fig7} represents the case when $g=1.0$, which is as well the onset of the bound and antibound state derived from the first-order strong-coupling perturbation theory (scpt). Comparing this to the numerical results in the non-adiabatic regime, we see that for a fixed $\omega_0$, the emergence of the bound state is shifted to higher values of the e-ph coupling while the emergence of the antibound state occurs always at $g<1.0$.

\section{Strong-coupling perturbation theory}

The phonon spectral function can be calculated analytically in some limiting cases. Starting from the SC limit $t=0$ where the Hamiltonian of Eq.~\ref{hol} describes a displaced harmonic oscillator on the site of the electron, we derive the expression for $B_q(\omega)$ in the first order perturbation of hopping $t$. We follow the procedure initiated in Ref.~\cite{bonca1,marsiglio95}.
Our aim is to show that both bound and antibound state can be described already by the 1st order scpt and their spectral weight in $B_q(\omega)$ compared to the numerical result.

Applying the canonical transformation $\tilde{H}=e^SHe^{-S}$ with $S=-g\sum_j (a_{j} - a_{j}^\dagger) n_j$, one gets the Hamiltonian $\tilde{H} = \tilde{H}_0 + \tilde{V}$, where the unperturbed part of $\tilde{H}$ is
\begin{equation} \label{h0}
\vspace*{-0.0cm}
\tilde{H}_0 = \omega_0\sum_{j} a_{j}^\dagger a_{j} - \epsilon_p  \sum_{j} n_{j},
\end{equation}
and $\tilde{V}$ is considered a perturbation
\begin{equation} \label{V}
\vspace*{-0.0cm}
\tilde{V} = -t e^{-g^2} \sum_{j} (c^\dagger_{j} c_{j+1} e^{-g(a_{j}^\dagger - a_{j+1}^\dagger)} e^{g(a_{j} - a_{j+1})}  +\mathrm{H.c.}).
\end{equation}
We have introduced the new basis where a harmonic oscillator has a shifted origin on the electron site $j$. We denote such state as $\vert g_j \rangle$. The ground state wave function at arbitrary $q$-point, which corresponds to the energy $E_q^0=-\epsilon_p-2te^{-g^2}\cos{q}$, can be written in translationally invariant form as $\vert\psi_q^0\rangle = 1/\sqrt{N}\sum_j e^{iqj} \vert g_j \rangle$, where $N$ is the number of lattice sites.
The bosonic operators $a_j$ and $a_{j}^\dagger$ in Eq.~\ref{h0} and~\ref{V} act relative to $\vert g_i \rangle$ if $j=i$, and relative to an unshifted oscillator if $j\neq i$.

\begin{figure}[!htb]
\includegraphics[width=1.0\columnwidth,clip]{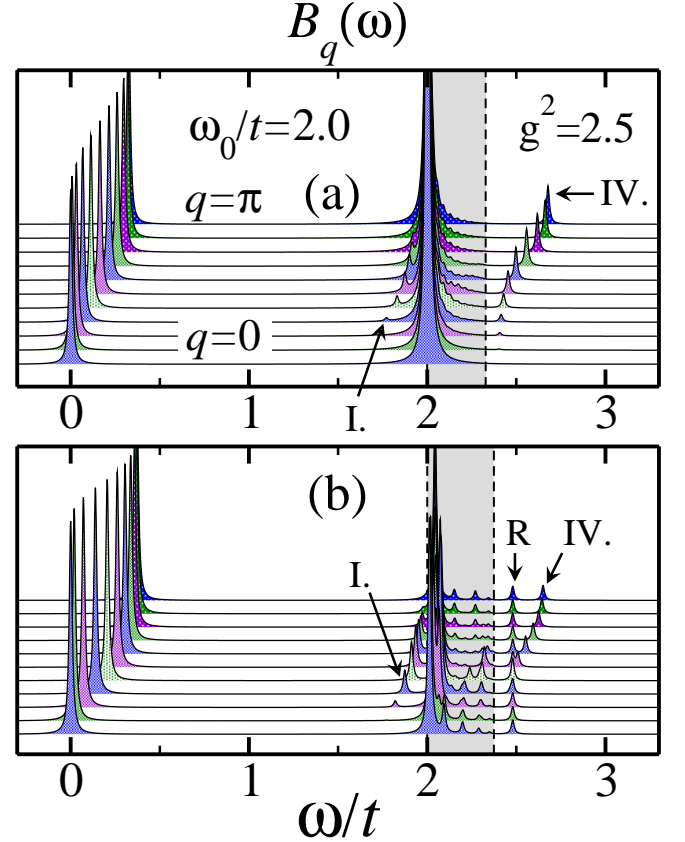}
\caption{(Color online) $B_q(\omega)$ for $\omega_0/t=2.0$ and $g^2=2.5$. (a) $B_q(\omega) = B_q^0(\omega) + B_q^1(\omega)$, calculated from the first-order scpt for $N=51$ sites (Eq.~\ref{b0} and Eq.~\ref{b1}). (b) Numerical results, obtained using the parameters $(N_h,M)=(8,11)$.
The bound peak (the Roman numeral $\mbox{I.}$) and the antibound peak (the Roman numeral $\mbox{IV.}$) can be observed in both cases. The repulsive peak (denoted by R) is obtained only in (b).
An artificial damping of $\epsilon=0.01$ was used.
}\label{fig8}
\end{figure}

Once the ground state wave function is known, the polaron spectral weight in $B_q(\omega)$ can be obtained from the matrix element
\begin{equation} \label{mat0}
\vspace*{-0.0cm}
\langle \psi_q^0 \vert a_q^\dagger + a_{-q} \vert \psi^0 \rangle = \frac{2g}{\sqrt{N}},
\end{equation}
which leads to the polaron peak
\begin{equation} \label{b0}
\vspace*{-0.0cm}
B_q^0(\omega) = \frac{4g^2}{N} \delta\left(\omega - 2te^{-g^2}(1-\cos{q})\right).
\end{equation}
The weight of the peak is proportional to the e-ph coupling parameter $g^2$. In Fig.~\ref{fig8}, $B_q(\omega)$ is shown at $\omega_0/t=2.0$ and $g^2=2.5$, calculated (a) from the first-order scpt and (b) numerically. The dispersion and the bandwidth of the polaron peak show good agreement since $\zeta$ yields $0.16$ in (a) and $0.18$ in (b).

The main focus of the calculation is the low-energy excitation spectrum. There are $N$ degenerate wave functions that consist of the translationally invariant electron state and an additional phonon quantum at a fixed distance $d$ away from the electron
\begin{equation} \label{psi1}
\vspace*{-0.0cm}
\vert\psi_d^1\rangle = \frac{1}{\sqrt{N}}\sum_j e^{iqj} a_{j+d}^\dagger \vert g_j \rangle.
\end{equation}
The definition of the bosonic operators is the same as in Eq.~\ref{h0} and~\ref{V}. To obtain the energies of the excited states one needs to calculate the matrix elements $\tilde{V}_{d,d'}^1 = \langle \psi_d^1 \vert \tilde{V} \vert\psi_{d'}^1\rangle$.
For $g>1$, there are $N-2$ solutions lying in the energy interval $[E_-^1,E_+^1]$, where $E_{\pm}^1=-\epsilon_p+\omega_0\pm 2te^{-g^2}$, and two additional states lying below and above the interval. These are the bound and antibound state, respectively. The emergence of these two states at $g=1$ is plotted by the dashed curve in Fig.~\ref{fig7}.
For $g\gtrsim 1$, the bound state first detaches from the continuum at $q=0$ and the antibound state at $q=\pi$.
When $g$ is further increased as shown in Fig.~\ref{fig8}(a) for $g^2=2.5$, these two states detach from the continuum for all values of $q$.
In the $g\to\infty$ limit, both bound and antibound state approach the energy $-\epsilon_p+\omega_0$.

The spectral weight of any one extra phonon excitation in $B_q(\omega)$ can be obtained from the matrix element
\begin{equation} \label{mat1}
\vspace*{-0.0cm}
\langle \psi_q^1 \vert a_q^\dagger \vert \psi^0 \rangle = \frac{1}{\sqrt{N}} \sum_d b_d^*(q) e^{iqd},
\end{equation}
where $\vert\psi_q^1\rangle = \sum_d b_d(q) \vert\psi_d^1\rangle$ is the corresponding eigenstate of the matrix $\tilde{V}^1$. Note that $\langle \psi_q^1 \vert a_{-q} \vert \psi^0 \rangle = 0$ for any state.
The total spectrum of the one extra phonon excitations is than
\begin{equation} \label{b1}
\vspace*{-0.0cm}
B_q^1(\omega) = \sum_i \frac{1}{N} \left| \sum_d b_d^{*(i)}(q) e^{iqd} \right|^2 \delta(\omega - (E_q^i-E^0)).
\end{equation}

The comparison of the phonon spectral function between the first-order scpt and numerical results is shown in Fig.~\ref{fig8} for $g^2=2.5$ and $\omega_0/t=2.0$. 
While the repulsive peak does not appear in the first-order scpt approximation,
the bound and antibound peak can be resolved from both spectra at very similar energy positions and comparable spectral weights.
The wave function of the bound state is symmetric at $q=0$ and antisymmetric at $q=\pi$, while the opposite is true for the antibound state.
This indicates that the essential physics of these two novel states is well captured by the first-order scpt, where the calculation of the one extra phonon excitations, as determined by the matrix elements $\tilde{V}_{d,d'}^1$ is analogous to the 1D tight-binding problem, modified around the position of the electron~\cite{bonca1}. In fact, if one assumes a regular tight-binding model with a non-zero onsite energy $e_0$ at only one site, its analytical solution yields an additional localized state below (for $e_0<0$) or above (for $e_0>0$) the continuous band of itinerant solutions.

\section{Summary}

In conclusion, the main focus of the paper has been to calculate the low-energy features of the phonon spectral function for the phonon energies $\omega_0$ ranging from adiabatic to antiadiabatic regime. Using an improved exact diagonalization method defined over a variational Hilbert space, we were able to get converged results in a wide range of the e-ph coupling strengths with special emphasis on the properties in the IC regime. In particular, we were interested in the low-energy excitation spectrum of the polaron, where the most spectral weight originates from the states where the polaron and the additional phonon excitation are unbound. These states are limited to the energy interval with the width equal to the polaron bandwidth $W$. Even though in our calculations we use a finite Hilbert space, the results are consistent with the existence of a continuum band of unbound excited states in the thermodynamic limit.

In addition, we have identified the emergence of novel states in the spectrum of the low-energy excitations for different regimes of model parameters.
Below the continuum band we have found three bound states of the polaron and additional phonon excitations, with the energy less than $\omega_0$ above the polaron ground state.
On the other hand, we have found one state above the one-phonon continuum denoted as the antibound state which consists of the polaron and additional phonon excitations with the energy higher than $\omega_0+W$ above the polaron ground state.
In both cases we showed that the spatial correlation between the polaron and extra phonon excitations decrease exponentially with the distance from the polaron. Consequently, the numerical method can calculate their properties to a sufficient accuracy.
Beside the bound and antibound states, we identified the state with a considerable spectral weight where the additional phonon excitation is repelled from the polaron. Due to the finite Hilbert space in our calculations, the energy of such repulsive state is always higher than that of the unbound states. However, we anticipate in the thermodynamic limit that this peak merges with the upper edge of the continuum band.

The bound and antibound states develop in the IC regime of the e-ph coupling and persist up to the SC limit. 
We have performed a detailed investigation of the properties of the antibound state, especially the onset of the corresponding peak in the phonon spectral function for a wide range of the model parameters, which lead to an alternative interpretation of this peak regarding its former explanation.
The antibound peak can be observed for the non-adiabatic values of $\omega_0$, whereas the adiabatic limit of the states above the $\omega_0+W$ threshold is not analyzed in the paper.
Our results can be compared to the analytical solution of the first-order scpt, valid for the antiadiabatic values of the phonon energies. In this approximation, the bound and antibound state can be observed for $\tilde{g}>\omega_0$. Nevertheless, the accurate numerical investigation yields even a lower value of the e-ph coupling for the onset of the antibound state (Fig.~\ref{fig7}), indicating that the antibound peak can be already observed before the cross-over to the small polaron regime.

\acknowledgments 

We acknowledge stimulating discussions with O. S. Bari\v si\'{c}, T. Tohyama and A. Ram\v sak.
J. B. acknowledge financial support of the SRA under grant P1-0044.

\bibliography{manuphhol}

\end{document}